\documentclass[12pt]{article}
 \usepackage{epsfig}
 \def\be{\begin{equation}}
 \def\ee{\end{equation}}
 \def\bea{\begin{eqnarray}}
 \def\eea{\end{eqnarray}}
 \usepackage{graphicx}

 \catcode`\@=11
 \def\lsim{\mathrel{\mathpalette\@versim<}}
 \def\gsim{\mathrel{\mathpalette\@versim>}}
 \def\@versim#1#2{\vcenter{\offinterlineskip
 \ialign{$\m@th#1\hfil##\hfil$\crcr#2\crcr\sim\crcr } }}
 \catcode`\@=12

 \catcode`@=12
\begin{document}
 \thispagestyle{empty}
 \begin{flushright}
 UCRHEP-T608\\
 Jan 2021\
 \end{flushright}
 \vspace{0.6in}
 \begin{center}
 {\LARGE \bf Linkage of Dirac Neutrinos to\\ 
 Dark U(1) Gauge Symmetry\\}
 \vspace{1.2in}
 {\bf Ernest Ma\\}
 \vspace{0.2in}
{\sl Physics and Astronomy Department,\\ 
University of California, Riverside, California 92521, USA\\}
\end{center}
 \vspace{1.2in}

\begin{abstract}\
It is shown how a mechanism which allows naturally small Dirac neutrino masses 
is linked to the existence of dark matter through an anomaly-free U(1) gauge 
symmetry of fermion singlets.
\end{abstract}

 \newpage
 \baselineskip 24pt
\noindent \underline{\it Introduction}~:~
There is a known mechanism since 2001~\cite{m01} for obtaining small Dirac 
fermion masses.  It was originally used~\cite{m01} in conjucntion with the 
seesaw mechanism for small Majorana neutrino masses, and later generalized 
in 2009~\cite{glr09}.  It has also been applied in 2016~\cite{m16} to light 
quark and lepton masses.  

The idea is very simple.  Start with the standard model (SM) of quarks and 
leptons with just one Higgs scalar doublet $\Phi = (\phi^+,\phi^0)$.  
Add a second Higgs scalar doublet $\eta = (\eta^+,\eta^0)$ which is 
distinguished from $\Phi$ by a symmetry yet to be chosen.  Depending on 
how quarks and leptons transform under this new symmetry, $\Phi$ and $\eta$ 
may couple to different combinations of fermion doublets and singlets. 
These Yukawa couplings are dimension-four terms of the Lagrangian which 
must obey this new symmetry.

In the Higgs sector, this new symmetry is allowed to be broken softly or 
spontaneously, such that $\langle \eta^0 \rangle = v'$ is naturally much 
smaller than $\langle \phi^0 \rangle = v$.  The mechanism is an analog 
of the well-known Type II seesaw for neutrino mass.  Consider for example the 
case where the new symmetry is global U(1) which is broken softly.  Let
\begin{eqnarray}
V &=& m_1^2 \Phi^\dagger \Phi + m_2^2 \eta^\dagger \eta + [\mu^2 \Phi^\dagger 
\eta + H.c.] \nonumber \\ 
&+& {1 \over 2} \lambda_1 (\Phi^\dagger \Phi)^2 + {1 \over 2} \lambda_2 
(\eta^\dagger \eta)^2 + \lambda_3 (\Phi^\dagger \Phi)(\eta^\dagger \eta) + 
\lambda_4 (\Phi^\dagger \eta)(\eta^\dagger \Phi),
\end{eqnarray}
where $\mu^2$ is the soft symmetry breaking term, then $v,v'$ are determined by
\begin{eqnarray}
0 &=& v[m_1^2 + \lambda_1 v^2 + (\lambda_3+\lambda_4) {v'}^2] + \mu^2 v', \\ 
0 &=& v'[m_2^2 + \lambda_2 {v'}^2 + (\lambda_3+\lambda_4) v^2] + \mu^2 v. 
\end{eqnarray}
For $m_1^2 < 0$ but $m_2^2 > 0$ and $|\mu^2| << m_2^2$, the solutions are
\begin{equation}
v^2 \simeq {-m_1^2 \over \lambda_1}, ~~~ v' \simeq {-\mu^2 v \over m_2^2 + 
(\lambda_3 + \lambda_4)v^2},
\end{equation}
implying thus $|v'| << |v|$.  In Ref.~\cite{m01}, the new symmetry is taken 
to be lepton number, under which $\eta$ has $L=-1$ but $\nu_R$ has $L=0$.  
This choice forbids $\bar{\nu}_R (\nu_L \phi^0 - l_L^- \phi^+)$, but allows 
$\bar{\nu}_R (\nu_L \eta^0 - l_L^- \eta^+)$.  Hence $\nu_L$ pairs up with 
$\nu_R$ to have a small Dirac mass through $v'$, but $\nu_R$ itself is 
unprotected by any symmetry so it may have a large Majorana mass $M$.  The 
end result is again a small Majorana neutrino mass proportional to ${v'}^2/M$. 
The difference is that $v'$ is naturally small already, so $M$ does not 
have to be much greater than the electroweak scale.

To explore further this mechanism, it is proposed that this new symmetry 
is gauged and that it enforces neutrinos to be Dirac fermions and requires 
the addition of neutral singlet fermions which become members of the dark 
sector, the lightest of which is the dark matter of the Universe. 

\noindent \underline{\it Dark U(1) Gauge Symmetry}~:~
The minimal particle content of the SM has only $\nu_L$, not $\nu_R$, and 
only the Higgs doublet $\Phi$.  Hence neutrinos are massless.  Knowing that 
they should be massive~\cite{pdg18}, the usuall remedy is to add $\nu_R$ 
and to assume that it pairs up with $\nu_L$ through $\phi^0$.  However, 
since $\nu_R$ is a particle outside the SM gauge framework, it has many 
possible different guises~\cite{m17}.  Here it will be assumed that it 
transforms under a new $U(1)_D$ gauge symmetry, whereas all SM particles 
do not.  The linkage of $\nu_R$ to the SM is achieved through a second 
Higgs doublet $\eta$ which transforms under $U(1)_D$ in the same way as 
$\nu_R$.  Using Eq.~(4) with a very large $m_2$, a sufficiently small 
$v'$, say of the order 1 eV, may be obtained for a realistic Dirac neutrino 
mass.

To be a legitimate and viable theory of Dirac neutrinos in this framework, 
there are two important conditions yet to be discussed.  First, the gauge 
$U(1)_D$ symmetry must not be broken in such a way that $\nu_R$ gets a 
Majorana mass.  Second, there must be additional fermions so that the 
theory is free of anomalies.  The two conditions are also connected 
because the additional fermions themselves must also acquire mass through 
the scalars which break $U(1)_D$.

Since only the new fermions transform under $U(1)_D$, the two conditions 
for anomaly freedom are
\begin{equation}
\sum_{i=1}^N n_i = 0, ~~~ \sum_{i=1}^N n_i^3 = 0,
\end{equation}
comprising of $N$ singlets, with $N$ to be determined.  There are some 
simple solutions:
\begin{itemize}
\item{(A) $(3,-2,-2,-2,-2,1,1,1,1,1)$.}
\item{(B) $(4,-3,-3,-3,2,2,1)$.}
\item{(C) $(5,-4,-4,1,1,1)$.}
\item{(D) $(6,-5,-5,3,2,-1)$.}
\end{itemize}
In the next two sections, solutions (B) and (C) will be examined in more 
detail because they allow both Dirac neutrinos and an associated dark 
sector in a consistent framework. Solution (D) will be mentioned briefly. 

\noindent \underline{\it Solution (B)}~:~
In addition to the seven singlet fermions listed, this scenario requires 
just the addition of $\eta \sim 1$ and a Higgs singlet $\chi \sim 1$ under 
$U(1)_D$.  Consider first the three fermion singlets $(4,2,2)$. They are not 
connected to one another because there are no scalar singlets transforming 
as 8,6,or 4. Consider then $(-3,-3,-3)$.  They are also not connected 
because $-6$ is missing.  However, $(4,2,2)$ is connected to $(-3,-3,-3)$ 
through $\chi \sim 1$ or $\chi^* \sim -1$.  This means that they form three 
massive Dirac fermions of magnitude determined by $\langle \chi \rangle = u$. 
As for the remaining singlet, it should be identified as $\nu_R \sim 1$. 
It pairs with $\nu_L$ through $\eta \sim 1$.  Because there is no scalar 
transforming as 2, $\nu_R$ does not get a Majorana mass.  It also cannot 
connect with the other singlets $(4,2,2)$ or $(-3,-3,-3)$.  This 
means that because of the chosen particle content of the model, there 
are two residual symmetries after the spontaneous breaking of $U(1)_D$, 
i.e. the usual lepton number and dark number under which $(4,2,2) \sim 1$ 
and $(-3,-3,-3) \sim -1$. 

The analog of Eq.~(1) is
\begin{eqnarray}
V &=& m_1^2 \Phi^\dagger \Phi + m_2^2 \eta^\dagger \eta + m_3^2 \chi^* \chi + 
[\mu \chi^* \Phi^\dagger \eta + H.c.] \nonumber \\ 
&+& {1 \over 2} \lambda_1 (\Phi^\dagger \Phi)^2 + {1 \over 2} \lambda_2 
(\eta^\dagger \eta)^2 + {1 \over 2} \lambda_3 (\chi^* \chi)^2 + \lambda_{12}
(\Phi^\dagger \Phi)(\eta^\dagger \eta) \nonumber \\ 
&+& \lambda'_{12} (\Phi^\dagger \eta)(\eta^\dagger \Phi) + \lambda_{13} 
(\Phi^\dagger \Phi)(\chi^* \chi) + \lambda_{23} (\eta^\dagger \eta) 
(\chi^* \chi).
\end{eqnarray}
The analog of Eq.~(4) is then
\begin{equation}
v^2 \simeq {-\lambda_3 m_1^2 + \lambda_{13} m_3^2 \over \lambda_1 \lambda_3 - 
\lambda_{13}^2}, ~~~ 
u^2 \simeq {-\lambda_1 m_3^2 + \lambda_{13} m_1^2 \over \lambda_1 \lambda_3 - 
\lambda_{13}^2}, ~~~ 
v' \simeq {-\mu u v \over m_2^2 + \lambda_{23} u^2 + 
(\lambda_{12}+\lambda'_{12}) v^2}.
\end{equation}
As an example, let $\mu \sim 10$ GeV, $u \sim 2$ TeV, $m_2 \sim 10^8$ GeV, 
then $v' \sim 0.5$ eV, which is of the order of neutrino masses.  
Since $m_2^2 > 0$ and large, $(\eta^+,\eta^0)$ are very heavy.  After the  
spontaneous breaking of $SU(2)_L \times U(1)_Y$ and $U(1)_D$, the only 
physical scalars are $h = \sqrt{2} Re(\phi^0)$ and $H = \sqrt{2} Re(\chi)$. 
They form the mass-squared matrix
\begin{equation}
{\cal M}^2_{hH} = \pmatrix{ 2\lambda_1 v^2 - \mu u v'/v & 2\lambda_{13} uv + 
\mu v' \cr 2\lambda_{13} uv + \mu v' & 2\lambda_3 u^2 - \mu v v'/u}.
\end{equation}
Assuming that $u^2 >> v^2$, the $h-H$ mixing is 
$(\lambda_{13}/\lambda_3)(v/u)$.

Let the Dirac fermion $\psi$ be the lightest linear combination of the 
$(4,2,2)$ and $(-3,-3,-3)$ dark fermions, with coupling to $H$ given by 
$f H \bar{\psi} \psi$, implying thus $m_\psi = \sqrt{2} f u$.  Its coupling 
to the $U(1)_D$ gauge boson $Z_D$ is assumed to be 
$g_D Z_D^\mu \psi \gamma_\mu \psi$.  Let $Z_D$ be lighter than $\psi$, 
then the relic abundance of $\psi$ is determined by its annihilation to 
$Z_D$, as shown in Fig.~1.
\begin{figure}[htb]
\vspace*{-5cm}
\hspace*{-3cm}
\includegraphics[scale=1.0]{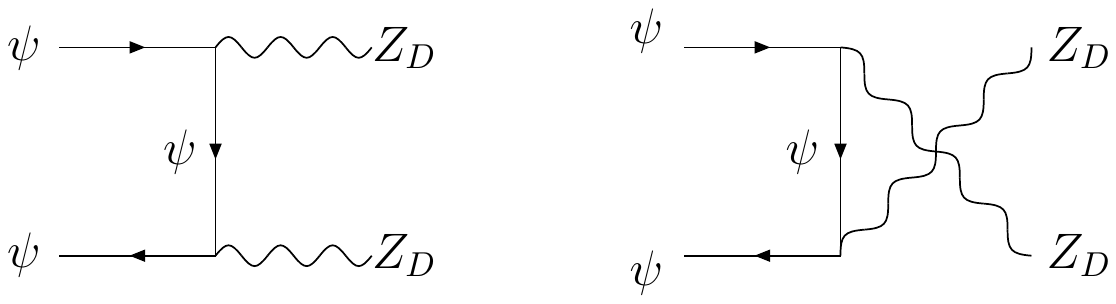}
\vspace*{-21.5cm}
\caption{Annihilation of $\psi \bar{\psi} \to Z_D Z_D$.}
\end{figure}
This cross section $\times$ relative velocity is given by
\begin{equation}
\sigma v_{rel} = {g_D^4 \over 16 \pi m_\psi^2} (1-m_D^2/m^2_\psi)^{3/2} 
(1-m_D^2/2m_\psi^2)^{-2}.
\end{equation}
Setting this value to the canonical $6 \times 10^{-26}~{\rm cm}^3/{\rm s}$ 
for a Dirac fermion, and assuming $m_\psi = 1$ TeV and $m_D = 800$ GeV, it 
is satisfied for $g_D = 0.86$, implying $u \simeq 2$ TeV.  Once produced, 
$Z_D$ decays quickly to two neutrinos.

As for the direct detection of $\psi$, it cannot proceed through $Z_D$ 
because the latter does not couple to quarks or charged leptons.  It may 
proceed through $h-H$ mixing.  For $m_\psi = 1$ TeV, the spin-independent 
cross section of dark matter scattering off a xenon nucleus is bounded 
by~\cite{xenon18} $10^{-45}$ cm$^2$.  This puts an upper limit of 
$4.55 \times 10^{-2}$ on $\lambda_{13}/\lambda_3$ for $u = 2$ TeV.

\noindent \underline{\it Solution (C)}~:~
The charge assignments of this scenario are well-known because $(5,-4,-4)$ 
is identical to $(-1,-1,-1)$ in terms of $\sum n_i$ and $\sum n_i^3$. 
Hence the former has been used to replace the latter as $B-L$ for three 
families of quarks and leptons, so that $B-L$ remains anomaly-free. 
It was first pointed out in 2009~\cite{mp09} and became the topic of some  
recent studies~\cite{ms15,cryz19,bccps20}.  Here they refer only to 
$U(1)_D$ under which the SM fermions do not transform.

The scalars required for this solution are a second doublet $\eta \sim -4$, 
and two singlets $\chi_1 \sim 2$ and $\chi_2 \sim 6$.  The $(-4,-4)$ fermions 
are identified as $\nu_R$, so they obtain Dirac masses through $\eta$, again 
with small $\langle \eta^0 \rangle = v'$.  The analog of Eq.~(6) is
\begin{eqnarray}
V &=& m_1^2 \Phi^\dagger \Phi + m_2^2 \eta^\dagger \eta + m_3^2 \chi_1^* 
\chi_1 + m_4^2 \chi_2^* \chi_2 \nonumber \\  
&+& [f_1 \chi_1^2 \Phi^\dagger \eta + f_2 \chi_2 \chi_1^* \Phi^\dagger 
\eta + f_3 \chi_2^* \chi_1^3 + H.c.] \nonumber \\ &+& {1 \over 2} \lambda_1 
(\Phi^\dagger \Phi)^2 + {1 \over 2} \lambda_2 (\eta^\dagger \eta)^2 + 
{1 \over 2} \lambda_3 (\chi_1^* \chi_1)^2 + {1 \over 2}\lambda_4 
(\chi_2^* \chi_2)^2 \nonumber \\ &+& \lambda_{12}(\Phi^\dagger \Phi)
(\eta^\dagger \eta) + \lambda'_{12} (\Phi^\dagger \eta)(\eta^\dagger \Phi) + 
\lambda_{13} (\Phi^\dagger \Phi)(\chi_1^* \chi_1) + \lambda_{14} 
(\Phi^\dagger \Phi)(\chi_2^* \chi_2) \nonumber \\ 
&+& \lambda_{23} (\eta^\dagger \eta) (\chi_1^* \chi_1) + \lambda_{24} 
(\eta^\dagger \eta) (\chi_2^* \chi_2) + \lambda_{34} (\chi_1^* \chi_1)
(\chi_2^* \chi_2).
\end{eqnarray}
Hence
\begin{equation}
v' = {-(f_1 u_1 + f_2 u_2)u_1 v \over m_2^2 + \lambda_{23} u_1^2 + 
\lambda_{24} u_2^2 + (\lambda_{12} + \lambda'_{12}) v^2}
\end{equation}
is again suppressed for large $m_2^2 >0$.

The $4 \times 4$ fermion mass matrix spanning $(5,1,1,1)$ has 9 nonzero 
entries from $u_1$ and 6 from $u_2$.  It is entirely disjoint from 
$(-4,-4)$.  This means that its lightest Majorana mass eigenstate $\zeta$ 
is a dark-matter candidate, stabilized with an odd dark parity.  After the 
spontaneous breaking of $SU(2)_L \times U(1)_Y$ and $U(1)_D$, excepting 
the heavy $\eta$, there are the $h$ and $H$ scalars as in Solution (B) 
as well as another scalar $H'$ and a pseudoscalar $A$, corresponding to 
the linear combination of $(u_2 \chi_1 - u_1 \chi_2)/\sqrt{u_1^2+u_2^2}$. 

Let $s = \sin \theta = u_1/\sqrt{u_1^2+u_2^2}$ and 
$c = \cos \theta = u_2/\sqrt{u_1^2+u_2^2}$, with  
\begin{equation}
\lambda'_3 = s^4 \lambda_3 + c^4 \lambda_4 + 2s^2c^2 \lambda_{34} + 
4s^3c f_3, ~~~ \lambda'_{13} = s^2 \lambda_{13} + c^2 \lambda_{14}.
\end{equation}
Then the analog of Eq.~(8) is
\begin{equation}
{\cal M}^2_{hH} = \pmatrix{ 2\lambda_1 v^2 & 2\lambda'_{13} uv  
\cr 2\lambda'_{13} uv & 2\lambda'_3 u^2}.
\end{equation}
The $h-H$ mixing is $(\lambda'_{13}/\lambda'_3)(v/u)$ and the trilinear 
$H^3$ coupling is $(\lambda'_3/\sqrt{2})u H^3$.

Assuming that $H$ is lighter than the dark-matter Majorana fermion $\zeta$, 
the annihilation of $\zeta \zeta \to HH$ is shown in Fig.~2.
\begin{figure}[htb]
\vspace*{-5cm}
\hspace*{-3cm}
\includegraphics[scale=1.0]{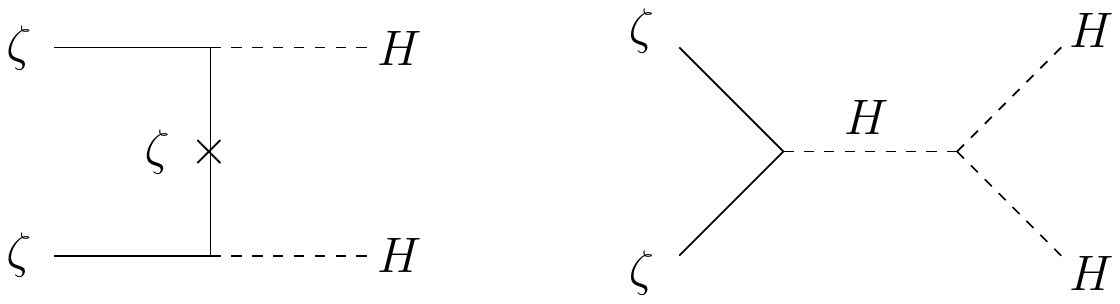}
\vspace*{-21.5cm}
\caption{Annihilation of $\zeta \zeta \to HH$.}
\end{figure}
The first diagram is also accompanied by its $u-$channel counterpart, which 
has the same amplitude in the limit that $\zeta$ is at rest.  Let 
$m_\zeta = f u \sqrt{2}$ and $x = m_H/m_\zeta$, then this cross section at 
rest multiplied by relative velocity is
\begin{equation}
\sigma \times v_{rel} = {f^4 \sqrt{1-x^2} \over 128 \pi m_\zeta^2} 
\left[ {2 \over 2-x^2} - {3x^2 \over 4-x^2} \right]^2.
\end{equation}
As an example, let $m_\zeta = 1$ TeV and $m_H = 400$ GeV, then the canonical 
value of $3 \times 10^{-26}~{\rm cm}^3/{\rm s}$ is obtained for $f=1.05$. 
This implies $u = 673$ GeV.  The limit on $\lambda'_{13}/\lambda'_3$ is 
then $2.7 \times 10^{-3}$ from XENON data~\cite{xenon18}.  In this 
scenario, the $Z_D$ gauge boson has $m^2_D = 8g_D^2 (s^2 + 9c^2)u^2$, 
so $m_D$ is of order a few TeV.

\noindent \underline{\it Solution (D)}~:~
Just as in (B), only one new Higgs doublet $\eta \sim -5$ and one singlet 
$\chi \sim -5$ are required.  The $(-5,-5)$ fermions are identified as 
$\nu_R$.  The $(6,-1)$ and $(3,2)$ pairs obtain independent masses from 
$\chi$, so there are two dark-matter components with two stabilizing 
symmetries.

\noindent \underline{\it Concluding Remarks}~:~
There are two often raised theoretical objections to having a Dirac neutrino.
(1) The singlet right-handed neutrino $\nu_R$ is trivial under the SM gauge 
symmetry, so it should have a Majorana mass.  (2) If a symmetry is invoked 
to forbid (1), then there is still no explanation as to why the Dirac neutrino 
mass is so small.  Here the answers are (1) that there is indeed a symmetry, 
i.e. a dark $U(1)_D$ gauge symmetry for $\nu_R$ but not the other SM 
particles, and (2) a small $v'$ from a second Higgs doublet $\eta$ 
transforming as $\nu_R$ under $U(1)_D$ is the source of this small Dirac 
mass.  It is obtained naturally by a (Type II) seesaw mechanism first 
pointed out in Ref.~\cite{m01}.  To implement this idea that Dirac neutrino 
mass is linked to a dark $U(1)_D$ gauge symmetry, a set of singlet fermions 
is required so that the theory is free of anomalies.  The scalars which 
are used to break $U(1)_D$ must be such that all fermions acquire mass, 
and two residual symmetries must remain: one is the usual lepton number, the 
other is a stabilizing dark symmetry.

Three examples are presented.  In (B), one singlet fermion is identified as 
$\nu_R$ whereas the other six form three dark Dirac fermions.  The 
stabilizing symmetry is global U(1).  In (C), two singlet fermions are 
identified as $\nu_R$ whereas the other four are Majorana fermions. 
The dark symmetry is $Z_2$ parity.  In (B), the lightest dark Dirac 
fermion $\psi$ annihilates to the $U(1)_D$ gauge boson $Z_D$ to establish 
its relic abundance.  In (C), it is the lightest dark Majorana fermion 
$\zeta$ annihilating to the $U(1)_D$ breaking scalar $H$.  In both cases, 
direct-search constraints put an upper limit on $h-H$ mixing of order 
$10^{-4}$.  In (D), dark matter consists of two separate Dirac fermion 
components.

\noindent \underline{\it Acknowledgement}~:~
This work was supported in part by the U.~S.~Department of Energy Grant 
No. DE-SC0008541.

\baselineskip 20pt

\bibliographystyle{unsrt}

\end{document}